\documentclass[aps,prb,twocolumn,showpacs]{revtex4}
\usepackage{graphicx}

\begin{document}

\title{Low temperature specific heat of the hole-doped Ba$_{0.6}$K$_{0.4}$Fe$_2$As$_2$ single crystals and electron-doped SmFeAsO$_{0.9}$F$_{0.1}$ samples}

\author{Gang Mu, Huiqian Luo, Zhaosheng Wang, Zhian Ren, Lei Shan, Cong Ren, and Hai-Hu Wen}\email{hhwen@aphy.iphy.ac.cn }

\affiliation{National Laboratory for Superconductivity, Institute of
Physics and Beijing National Laboratory for Condensed Matter
Physics, Chinese Academy of Sciences, P.O. Box 603, Beijing 100190,
People's Republic of China}

\begin{abstract}
Low temperature specific heat (SH) was measured on the
 FeAs-based superconducting single crystals
Ba$_{0.6}$K$_{0.4}$Fe$_2$As$_2$ and high pressure synthesized
polycrystalline samples SmFeAsO$_{0.9}$F$_{0.1}$. It is found that
the sharp SH anomaly $\Delta C/T|_{T_c}$ in
Ba$_{0.6}$K$_{0.4}$Fe$_2$As$_2$ reaches an unexpected high value of
98 mJ/mol K$^2$, about one order of magnitude larger than that of
SmFeAsO$_{0.9}$F$_{0.1}$ ($6\sim8$ mJ/mol K$^2$) samples, suggesting
very high normal state quasiparticle density of states in FeAs-122
than in FeAs-1111. Furthermore, we found that the electronic SH
coefficient $\gamma_e(T)$ of Ba$_{0.6}$K$_{0.4}$Fe$_2$As$_2$ is
weakly temperature dependent and increases almost linearly with the
magnetic field in low temperature region, which may indicate that
the hole-doped FeAs-122 system contains a dominant component with a
full superconducting gap, although we cannot rule out the
possibility of a small component with anisotropic or nodal gap. A
detailed analysis reveals that the $\gamma_e(T)$ of
Ba$_{0.6}$K$_{0.4}$Fe$_2$As$_2$ cannot be fitted with a single gap
of s-wave symmetry probably due to the multigap effect. These
results indicate clear difference between the properties of the
superconducting state of the holed-doped
Ba$_{0.6}$K$_{0.4}$Fe$_2$As$_2$ and the F-doped LnFeAsO (Ln = rare
earth elements) systems, which we believe is originated from the
complex Fermi surface structures in different systems.

\end{abstract} 
\maketitle

\section{Introduction}

The discovery of high temperature superconductivity in the
FeAs-based system has stimulated enormous interests in the fields of
condensed matter physics and material sciences \cite{Kamihara2008}.
The superconductivity has not only been discovered in the
electron-doped samples, but also in the hole-doped
ones\cite{Wen2008,Rotter1}. The central issues concerning the
superconductivity mechanism are about the pairing symmetry and the
magnitude of the superconducting gap. The experimental results
obtained so far are, however, highly controversial. The low
temperature specific heat (SH) measurements in the F-doped $LaFeAsO$
samples revealed a nonlinear magnetic field dependence of the SH
coefficient $\gamma_e$, which was attributed to the presence of a
nodal gap\cite{MuG}. This was later supported by many other
measurements based on $\mu$SR\cite{uSR1,uSR2,uSR3},
NMR\cite{NMR1,NMR2,NMR3}, magnetic penetration\cite{Hc1} and point
contact Andreev spectrum (PCAS)\cite{ShanL}. On the other hand, the
PCAS on the F-doped SmFeAsO indicated a feature of s-wave
gap\cite{Chien}, some measurements\cite{swave1,swave2,swave3,swave4}
also gave support to this conclusion. It is important to note that
most of the conclusions drawn for a nodal gap were obtained on the
electron-doped LnFeAsO samples (abbreviated as FeAs-1111, Ln stands
for the rare earth elements) which are characterized by a low charge
carrier density and thus low superfluid density\cite{ZhuXY}. For the
FeAs-1111 phase, it is very difficult to grow crystals with large
sizes, therefore most of the measurements on the pairing symmetry so
far were made on polycrystalline samples. This is much improved in
the (Ba,Sr)$_{1-x}$K$_x$Fe$_2$As$_2$ (denoted as FeAs-122) system
since sizable single crystals can be achieved\cite{LuoHQ}.
Preliminary data by angle resolved photo-emission spectroscopy
(ARPES) on these crystals show two groups of superconducting gaps
($\Delta_1\approx$ 12 meV, $\Delta_2 \approx$ 6 meV) all with s-wave
symmetry\cite{DingH,ZhouXJ,Hasan}. It is known that the surface of
this type of single crystals decay or reconstruct very quickly, this
may give obstacles to get repeatable data when using the surface
sensitive tools. Thus solid conclusions about the gap symmetry and
magnitude from bulk measurements are highly desired.

Specific heat (SH) is one of the powerful tools to measure the bulk
properties, especially about the quasiparticle density of states
(DOS) at the Fermi level. By measuring the variation of the
electronic SH versus temperature and magnetic field, one can
essentially determine the feature of the gap symmetry. In this
paper, for the first time, we report the detailed low temperature SH
data on Ba$_{0.6}$K$_{0.4}$Fe$_2$As$_2$ single crystals with $T_c$ =
36.5 K (90\%$\rho_n$). We also present the SH data on a high
pressure synthesized F-doped SmFeAsO sample for comparison. Our
results reveal the existence of a dominant component with a full
superconducting gap in Ba$_{0.6}$K$_{0.4}$Fe$_2$As$_2$, although we
could not rule out the possibility that there might be small
component of superfluid with the nodal gap. Therefore the FeAs-122
is very different from the case in the F-doped LnFeAsO where a nodal
gap feature was discovered. Meanwhile we show the evidence of an
unexpected large SH anomaly in the FeAs-122 superconductors, which
is associated with the large DOS in the normal state. These two
features indicate that there may be clear differences between the
hole-doped FeAs-122 and the electron-doped FeAs-1111 phases. We
argue that these differences are originated from the complex Fermi
surface structures in different systems.\cite{Zabolotnyy} In the
Ba$_{0.6}$K$_{0.4}$Fe$_2$As$_2$ samples, the outer shell of the FSs
surrounding $\Gamma$ point takes most of the weight and exhibits a
full gap. While in the F-doped LnFeAsO, the FSs near the M points
contribute more weight to the superfluid and they have radical
momentum dependence and nodes may be anticipated.

\begin{figure}
\includegraphics[width=8.5cm]{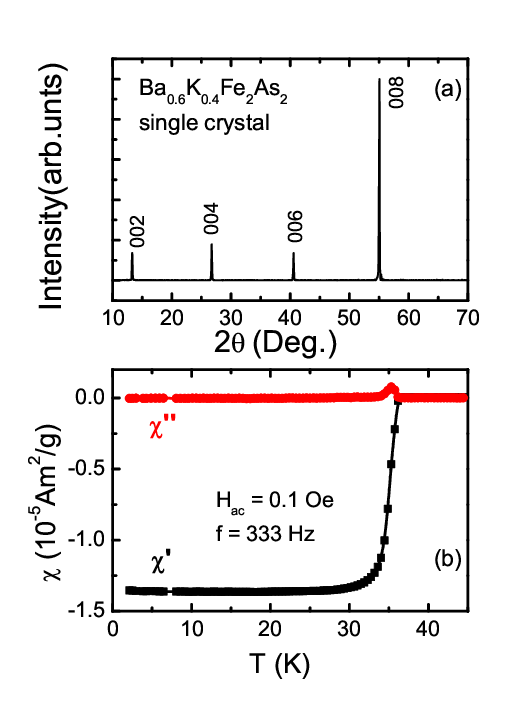}
\caption {(color online) (a) X-ray diffraction pattern measured for
the single crystal of Ba$_{0.6}$K$_{0.4}$Fe$_2$As$_2$ with $T_c$ =
36.5 K. Only sharp peaks along (00$l$) orientation were observed
with the full-width-at-half-maximum (FWHM) of about 0.10 $^\circ$.
(b) Temperature dependence of ac susceptibility for the same sample
measured with $H_{ac}$ = 1 Oe and $f$ = 333 Hz. One can see a rather
sharp transition with the onset transition temperature of about 36
K.} \label{fig1}
\end{figure}

\section{Sample Preparation and Characterization}

The superconducting single crystals of Ba$_{1-x}$K$_{x}$Fe$_2$As$_2$
with $T_c$ of about 36.5 K were grown by using FeAs as the self-flux
\cite{LuoHQ}. The weighing and mixing procedures were performed in a
glove box with a protective argon atmosphere. The mixed materials
were placed in an alumina oxide crucible and sealed under vacuum in
a quartz tube. The melting process was carried out at high
temperatures of 1000 $\sim$1150 $^o$C, and then a slowly cooling
process was followed. The sample we chose for the SH measurement has
the dimension of 3.0 $\times$ 1.5 $\times$ 0.2 mm$^{3}$. The
potassium content was estimated to be about 40\% from the EDX
analysis\cite{LuoHQ}.

The X-ray diffraction (XRD) of the single crystals was carried out
by a $Mac$-$Science$ MXP18A-HF equipment with $\theta - 2\theta$
scan. The ac susceptibility were measured based on an Oxford
cryogenic system (Maglab-Exa-12). The resistivity and the specific
heat were measured with a Quantum Design instrument physical
property measurement system (PPMS) with the temperature down to 1.8
K and the magnetic field up to 9 T. The temperature stabilization
was better than 0.1\% and the resolution of the voltmeter was better
than 10 nV. We employed the thermal relaxation technique to perform
the specific heat measurements. To improve the resolution, we used a
latest developed SH measuring frame from Quantum Design, which has
negligible field dependence of the sensor of the thermometer on the
chip as well as the thermal conductance of the thermal linking
wires.

\begin{figure}
\includegraphics[width=9cm]{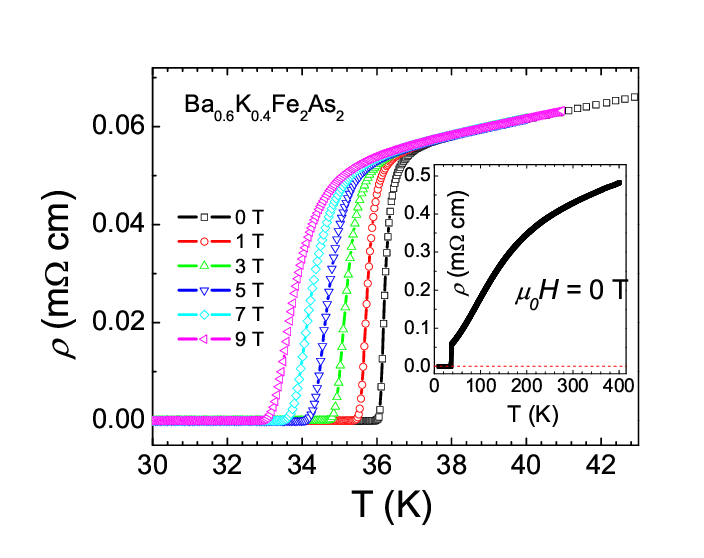}
\caption {(color online) Temperature dependence of the resistivity
near superconducting transition for the
Ba$_{0.6}$K$_{0.4}$Fe$_2$As$_2$ sample at magnetic fields ranging
from 0 to 9 T with H $\parallel$ c axis. The inset shows the
resistive curve under zero field up to 400 K.} \label{fig2}
\end{figure}

The crystal structure of sample Ba$_{0.6}$K$_{0.4}$Fe$_2$As$_2$ with
$T_c$ = 36.5 K was examined by XRD measurement with incident X-ray
along the $c$-axis direction. The XRD pattern is shown in Fig. 1(a).
It is clear that only sharp peaks along (00$l$) orientation can be
observed, suggesting a high $c$-axis orientation. The
full-width-at-half-maximum (FWHM) of the diffraction peaks were only
about 0.10 $^\circ$, indicating a rather fine crystalline quality in
the present sample. The diamagnetic transition measured with ac
susceptibility technique is shown in Fig. 1(b). A rather sharp
transition at about 36 K (onset) can be seen, which provides an
evidence for the homogeneity of the superconducting properties of
our sample.

In Fig. 2, we show the temperature dependence of the in-plane
electrical resistivity near $T_c$ at magnetic field up to 9 T with
$H\|c$ axis for the sample Ba$_{0.6}$K$_{0.4}$Fe$_2$As$_2$. The
onset transition temperature (90\%$\rho_n$) is determined to be
about 36.5 K under zero field. And a rather narrow transition width
($\Delta T_c$$\sim$1 K) can be observed when the magnetic field is
lower than 3 T, which also suggests the high quality of the present
sample. By applying a magnetic field the middle transition point
(50\%$\rho_n$) shifts to lower temperatures slowly with a slope $-d
\mu_0H_{c2}(T)/dT|_{T_c} \approx$ 4.1 T / K. Using the
Werthamer-Helfand-Hohenberg relation\cite{WHH}
$\mu_0H_{c2}(0)=-0.69d \mu_0H_{c2}(T)/dT|_{T_c} T_c$, we get the
upper critical field $\mu_0H_{c2}(0) \approx$ 100 T ($H\|c$). The
inset of Fig. 2 shows temperature dependence of the resistivity in a
wide temperature regime up to 400 K at zero field. One can see that
the $\rho-T$ curve exhibits a continued curvature in the normal
state up to 400 K.

The fluorine doped SmFeAsO superconducting samples were prepared by
a high pressure synthesis method. The details about the synthesis
were reported previously.\cite{RenZACPL} As shown in Fig. 3, the ac
susceptibility data (measured using an ac amplitude of 0.1 Oe)
exhibits a sharp magnetic transition. The width defined between the
10\% and 90\% cuts of the transition is below 2 K, with the middle
transition point at 51.5 K, indicating a very good quality of the
superconducting phase. The inset shows a scanning electron
microscope picture of the sample. Compared with the samples
synthesized at ambient pressures, the samples studied here are much
more compact. A rough estimate on the diamagnetic signal indicates
that the superconducting volume is close to 100 \%. The relatively
sharp superconducting transition and the large fraction of the
superconducting volume validate the specific heat measurements and
the data analysis in this paper.

\section{Analysis to the Specific heat data of $\texttt{Ba}_{0.6}\texttt{K}_{0.4}\texttt{Fe}_2\texttt{As}_2$}
\begin{figure}
\includegraphics[width=9cm]{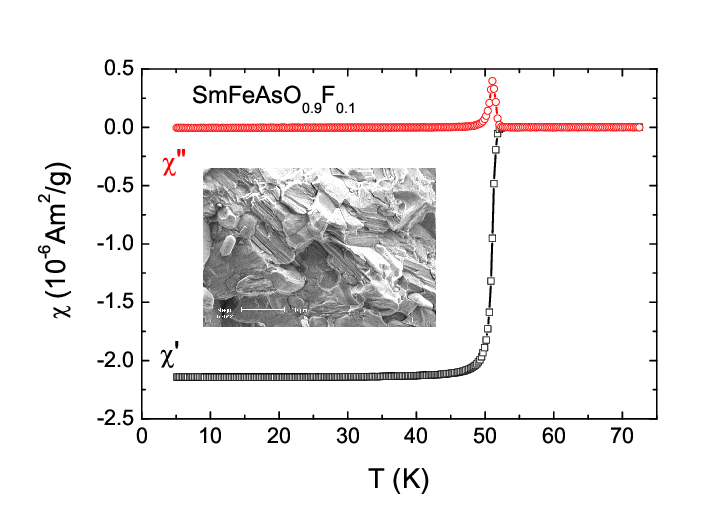}
\caption {(color online) Temperature dependence of the ac
susceptibility of the high pressure synthesized bulk sample
SmFeAsO$_{0.9}$F$_{0.1}$ measured with an ac field of 0.1 Oe and
frequency of 333 Hz. A very sharp superconducting transition is
obvious. A rough estimate on the magnetic signal indicates that the
superconducting volume is close to 100 \%. The inset shows a
scanning electron microscope picture of the sample. The grains with
plate-like shape are highly compacted together.} \label{fig3}
\end{figure}

\begin{figure}
\includegraphics[width=8.5cm]{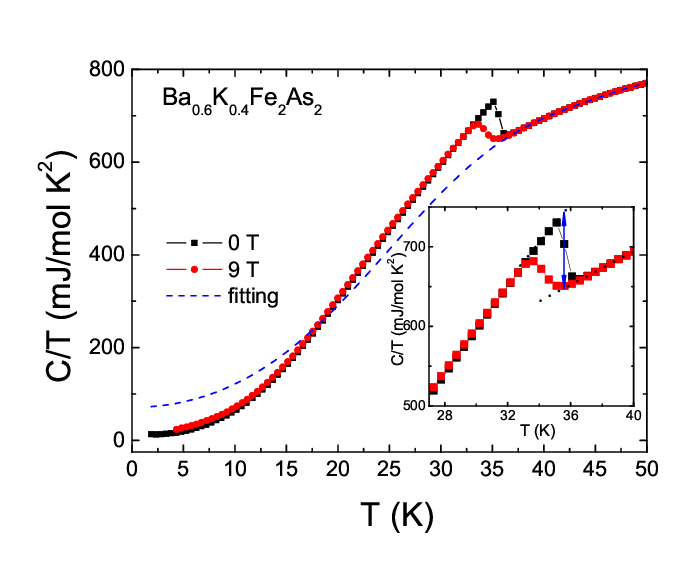}
\caption {(color online) Raw data of SH coefficient $\gamma=C/T$ vs
$T$ for Ba$_{0.6}$K$_{0.4}$Fe$_2$As$_2$ sample are shown in the main
frame. The dashed line shows the normal state SH obtained from
fitting to Eq. (6). The inset shows an enlarged view of the data
$\gamma=C/T$ near $T_c$. The sharp SH anomaly $\Delta C/T|_{T_c}$ is
indicated by the arrowed blue short line with a magnitude of about
98 mJ/mol K$^2$.} \label{fig4}
\end{figure}

\begin{figure}
\includegraphics[width=8.5cm]{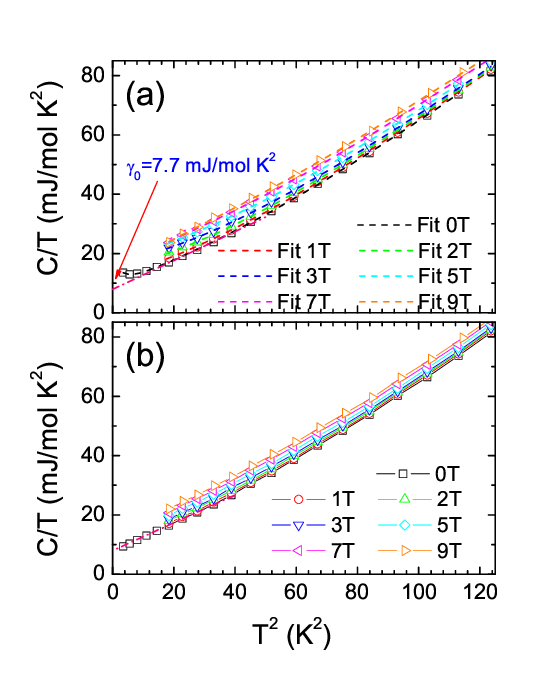}
\caption {(color online) Temperature and magnetic field dependence
of specific heat in $C/T$ vs $T^2$ plot in the low temperature
range. (a) Raw data before removing the Schottky anomaly. The dashed
lines represent the theoretical fit (see text) containing all terms
in Eq. (4). (b) Replot of the data after the Schottky anomaly was
subtracted. The dot-dashed line represents an extension of the zero
field data to T = 0 K giving a residual value $\gamma_0$ = 7.7
mJ/mol K$^2$(see text).} \label{fig5}
\end{figure}

In the main panel of Fig. 4 we show the raw data of SH coefficient
$\gamma=C/T$ vs $T$ for Ba$_{0.6}$K$_{0.4}$Fe$_2$As$_2$ sample at 0
T and 9 T. Multiple complicated contributions to the SH data emerged
in the low-T region when a magnetic field was applied, so we only
show the data above 4.3 K under magnetic fields, and the data at
zero field was shown down to about 1.8 K. Clear and sharp
superconducting anomalies can be seen near $T_c$ from the raw data.
The SH anomaly $\Delta C/T|_{T_c}$ at zero field was determined to
be about 98 mJ/mol K$^2$, indicated by the vertical short blue line
in the inset of Fig. 4. This is remarkably different from the case
in the FeAs-1111 phase, where only a very small SH anomaly (less
than 8 mJ/mol K$^2$) was observed in the raw
data\cite{MuG,David,SYLi}. In section V, we will display the SH data
of a high-pressure synthesized sample SmFeAsO$_{0.9}$F$_{0.1}$,
where the SH anomaly near $T_c$ reaches only 6$\sim$8 mJ/mol K$^2$.
Therefore the large value of $\Delta C/T|_{T_c}$ in
Ba$_{0.6}$K$_{0.4}$Fe$_2$As$_2$ is an intrinsic property, it may be
associated with the high quasiparticle DOS in the normal state,
which will be further addressed later. A magnetic field of 9 T
shifts the SH anomaly down for only 1.5 K and suppresses the anomaly
clearly. This is consistent very well with the resistive
measurements which indicate that the upper critical field is very
high in this system.\cite{WangZSPRB}Below about 10 K, a clear
flattening feature of $C/T$ can be seen in Fig. 4, which may imply
the weak excitation of quasiparticles.

The raw data in the low temperature region at different fields are
plotted as $C/T$ vs $T^2$ in Fig. 5(a). A slight curvature was
detected in the region from 4.3 K to 11 K on the plot of $C/T$ vs.
$T^2$, which was attributed to the electron contribution of the
superconducting state and the quintic term of the phonon
contribution. This brought in enormous difficulties when treating
the data because it gave too many fitting parameters. Therefore we
first analyze the data below 6 K at zero field, where the two terms
mentioned above remain negligible. Consequently the zero field data
below 6 K can be represented by the following equation:
\begin{equation}
C(T,H=0)=\gamma_0 T+\beta T^3+C_{Sch}(T,H=0),\label{eq:1}
\end{equation}
where the three terms represent the contributions of the residual
electronic SH, the phonon and the magnetic impurity (the so-called
Schottky anomaly), respectively. The Schottky anomaly is generally
induced by lifting the degeneracy of the states of the paramagnetic
spins in the magnetic impurities\cite{Sch,MgIrB} and is given by
\begin{equation}
C_{Sch}(T,H)=n(\frac{g\mu_B H_{eff}}{k_B T})^2 \frac{e^{\frac{g\mu_B
H_{eff}}{k_B T}}}{(1+e^{\frac{g\mu_B H_{eff}}{k_B
T}})^2},\label{eq:2}
\end{equation}
where g is the Land\'{e} factor, $\mu_B$ is the Bohr magneton,
H$_{eff}$=$\sqrt{H^2+H_{0}^{2}}$ is the effective magnetic field
which evolves into $H_{eff}$ = $H_0$, the crystal field at zero
external field, and $n$ is the concentration of paramagnetic
centers. The obtained fitting parameters $\gamma_0 \approx$ 7.7
mJ/mol K$^2$ and $\beta\approx$ 0.473 mJ/mol K$^4$ are very close to
the values obtained by simply drawing a linear line below 6 K as
shown by the dot-dashed line in Fig. 5(a). The values of $\gamma_0$
and $\beta$ are then fixed when fitting the zero field data up to 11
K, where all the terms must be taken into account:
\begin{equation}
C(T,H=0)=\gamma_0 T+[\beta T^3+\eta
T^5]+C_{es}+C_{Sch}(T,H=0),\label{eq:3}
\end{equation}
where $\eta$ is the quintic term coefficient of the phonon SH and
$C_{es}=D\times e^{-\Delta(0)/k_B T}/T^{1.5}$ is the superconducting
electron contribution based on an s-wave scenario, with $\Delta(0)$
the superconducting gap at 0 K.  By fitting the data at zero field
using Eq. (3), we obtained $\eta \approx$ 0.00034 mJ/mol K$^6$ and
$\Delta(0) \approx$ 5.99 $\pm$ 0.03 meV. As for the data under
finite fields, a magnetic field induced term $\gamma(H)$ arises and
the total SH can be written as
\begin{equation}
C(T,H)=[\gamma_0+\gamma(H)] T+[\beta T^3+\eta
T^5]+C_{es}+C_{Sch}(T,H).\label{eq:4}
\end{equation}
It is quite rational to fix $\gamma_0$, $\beta$, and $\eta$ as the
values obtained from analyzing the data at zero field. And the
superconducting gap under magnetic fields was restricted using the
relation\cite{Maki} $\Delta(H)=\Delta(0)\sqrt{1-H/H_{c2}}$ assuming
a field induced pair breaking effect. In this way the number of the
fitting parameters were reduced remarkably and creditable results
can be obtained.

We should however, note that in treating the data with above
equations, we assumed a dominant contribution of superfluid with
s-wave symmetry. This does not mean that we can rule out the
possibility of some small components of superfluid with a nodal gap.
This is because a component with nodes will contribute a power law
term to the specific heat in the low temperature limit, which cannot
be easily separated from the combination of the residual term
$\gamma_0T$ and the phonon term $\beta T^3$. However, from the field
dependence of the specific heat, we confirm that this possible small
component with nodes takes only a very small part of the total
condensate, since the specific heat coefficient increases almost
linearly with the magnetic field, as predicted for an s-wave gap.

Fig. 5(b) shows the data after removing the Schottky anomaly from
the total SH. The obtained field induced term $\gamma(H)$ are shown
in Fig. 6 (will be discussed later) and the fitting parameters
related to the terms $C_{es}$ and $C_{Sch}$ are shown in Table I.
The obtained residual term $\gamma_0 \approx$ 7.7 mJ/mol K$^2$
accounts for about 11\% of the total electron contribution (will be
discussed later), indicating a superconducting volume fraction of
about 89\% in our sample. Using the obtained value of $\beta$ and
the relation $\Theta_D$ = $(12\pi^4k_BN_AZ/5\beta)^{1/3}$, where
$N_A$ = 6.02 $\times 10^{23}$ mol$^{-1}$ is the Avogadro constant, Z
= 5 is the number of atoms in one unit cell, we get the Debye
temperature $\Theta_D \approx$ 274 K. This value is close to that
found in the LaFeAsO$_{0.9}$F$_{0.1-\delta}$ system\cite{MuG}.

\begin{table}
\caption{Fitting parameters ($\Delta$ is calculated through
$\Delta(H)=\Delta(0)\sqrt{1-H/H_{c2}}$, $H_0$ is fixed with the
value from the fitting to zero-field data).}
\begin{tabular}{cccccc}
\hline \hline
$\mu_0H$(T) & $D$(mJ$\,$K$^{0.5}/$mol) & $\Delta$(meV) & $n$(mJ$/$mol$\,$K) & $\mu_0H_0$(T)  & $g$ \\
\hline
0.0      & $2.11\times10^{6}$       & $5.99$      & $20.95$        & $1.70$      & $2.75$ \\
1.0      & $2.07\times10^{6}$       & $5.96$      & $31.08$        & $1.70$      & $3.18$ \\
2.0      & $1.93\times10^{6}$       & $5.93$      & $35.76$        & $1.70$      & $3.26$ \\
3.0      & $1.82\times10^{6}$       & $5.90$      & $36.22$        & $1.70$      & $3.38$ \\
5.0      & $1.63\times10^{6}$       & $5.84$      & $36.11$        & $1.70$      & $3.24$ \\
7.0      & $1.57\times10^{6}$       & $5.78$      & $36.71$        & $1.70$      & $3.11$ \\
9.0      & $1.43\times10^{6}$       & $5.72$      & $35.98$        & $1.70$      & $3.22$ \\
 \hline \hline
\end{tabular}
\label{tab:table1}
\end{table}

\section{A dominant component of superfluid with s-wave symmetry and very high DOS in $\texttt{Ba}_{0.6}\texttt{K}_{0.4}\texttt{Fe}_2\texttt{As}_2$}

The field-induced change of the electron SH coefficient $\gamma(H)$
was obtained from fitting and plotted in Fig. 6. It can be seen
clearly that $\gamma(H)$ increases almost linearly with the magnetic
field in the temperature region up to 11 K. A linear fit with the
slope of about 0.633 mJ/mol K$^2$ T to the zero temperature data is
revealed by the blue solid line in this figure. It is clear that the
$\gamma(H)-H$ curve roughly displays a linear behavior at all
temperatures below 11 K. This linear behavior is actually
anticipated by the theoretical prediction for superconductors with a
full gap\cite{Hussey}, in which $\gamma(H)$ is mainly contributed by
the localized quasiparticle DOS within vortex cores. So it seems
that the superconductivity in the FeAs-122 phase is primarily
dominated by a full-gap state around the hole-like Fermi surface at
$\Gamma$ point and the field-induced quasiparticle DOS are mainly
contributed by the vortex cores. This is in sharp contrast with the
results in cuprates\cite{Moler,Junod,Phillips,WenHH} and the
LaFeAsO$_{0.9}$F$_{0.1-\delta}$ system\cite{MuG} where a
$\gamma(H)\propto\sqrt{H}$ relation was observed and attributed to
the Doppler shift of the nodal quasiparticle spectrum. The curve
plotted using the relation $\gamma(H)$ = A$\sqrt{H}$ is also
presented in Fig. 6. It is obvious that this curve fails to fit our
data. We have also attempted to fit the data by mixing linear and
square root behavior as revealed by the navy-blue solid line in Fig.
6. One can see that this curve can also roughly describe the
behavior of our data. So there is the possibility for the presence
of a d-wave gap at different positions of the Fermi surface.
\begin{figure}
\includegraphics[width=9cm]{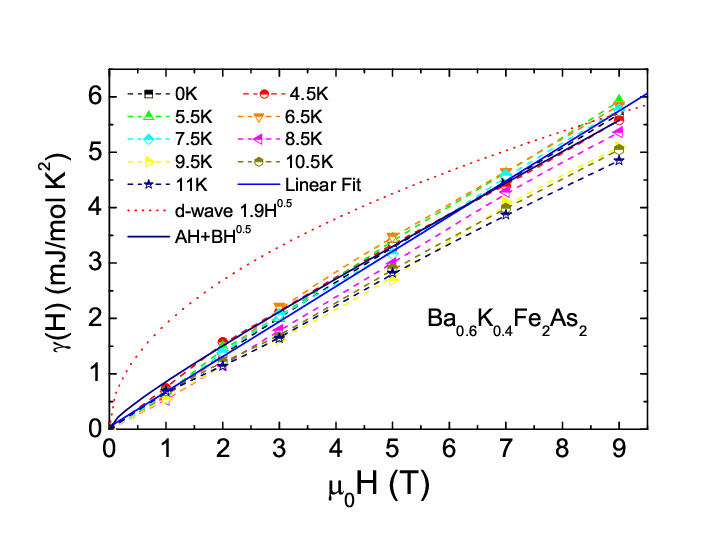}
\caption{(color online) Field dependence of the field-induced term
$\gamma(H)$ at temperatures ranging from 4.5 K to 11.1 K, and that
at $T = 0$ K obtained from fitting (see text). The dashed lines for
different temperatures are guides for the eyes. The blue solid, red
dotted, and navy-blue solid line are the linear fit to the zero
temperature data, the fit to the d-wave prediction $\gamma(H)$ =
A$\sqrt{H}$, and the fit by mixing the above two components,
respectively. } \label{fig6}
\end{figure}

In order to further confirm this point, we analyzed the SH data in
finite temperature region in the mixed state. It is known that the
quasiparticle excitations in superconductors with different gap
symmetry can be obviously distinct. In s-wave superconductors, the
inner-core states dominate the quasiparticle excitations, and
consequently a simple scaling law $C_{core}/T^3\approx
(\gamma_n/H_{c2(0)})\times(T/\sqrt{H})^{-2}$ for the fully gapped
superconductors is expected. While for a gap with line nodes, the
excitation spectrum is dominated by the extended quasiparticles
outside the vortex cores. And the so-called Simon-Lee scaling
law\cite{SimonLee} $C_{vol}/(T\sqrt{H})=f(T/\sqrt{H})$ should be
obeyed. A simple analysis similar to that has been done in our
previous work\cite{LiuZY} shows that for the superconductor with an
s-wave symmetry,
$C_{cal-s}=[(C(H)-C_{Sch}(H))-(C(H=0)-C_{Sch}(H=0))]/T^{3} \approx
C_{core}/T^{3}$, considering the fact that the phonon SH is
independent on the magnetic field. In other words, the defined term
$C_{cal-s}$ should scale with $T/\sqrt{H}$ with the prefactor
$\gamma_n/H_{c2(0)}$. Similarly for the d-wave symmetry we have
known\cite{LiuZY,WenHH} that
$C_{cal-d}=[(C(H)-C_{Sch}(H))-(C(H=0)-C_{Sch}(H=0))]/T\sqrt{H} =
C_{vol}/T\sqrt{H}-\alpha T/\sqrt{H}$, where $\alpha$ is the electron
SH coefficient at zero field for a d-wave superconductor, should
also scale with $T/\sqrt{H}$. The scaling result of the
field-induced term in the mixed state with the s-wave condition is
presented in Fig. 7(a). One can see that all the data at different
magnetic fields can be scaled roughly to one straight line, which
reflects the theoretical curve $C_{cal-s}= 0.633\times
(T/\sqrt{H})^{-2}$. Naturally, this prefactor $\gamma_n/H_{c2(0)}=
0.633$ mJ/(mol K$^2$ T) is consistent with the magnitude of the
slope of the blue line in Fig. 6. Using the value of $H_{c2(0)}
\approx 100$ T, we can estimate the normal state electron SH
coefficient $\gamma_n$ of about 63.3 mJ/mol K$^2$, a rather large
value compared with the F-doped LaFeAsO system\cite{MuG} and SmFeAsO
system which will be revealed in the next section. Fig. 7(b) shows
the scaling by following the d-wave scheme. It is clear that the
s-wave scaling is much batter than that of the d-wave case. This
again indicates that the superconducting state in this FeAs-112
phase is dominated by an s-wave symmetry gap and the field-induced
quasiparticle DOS are mainly contributed by the vortex cores. And
the slight departure of the scaling in Fig. 7(a) may suggest that
small amount of d-wave components exist in the superconducting
state.

Using the value of $\gamma_n\approx$ 63.3 mJ/mol K$^2$, we get the
ratio $\Delta C_e/\gamma_nT|_{T_c} \approx $ 1.55 being very close
to the weak-coupling BCS value 1.43. Considering the electron
re-normalization effect, the electron SH coefficient of a metal can
be written as
\begin{equation}
\gamma_n=\frac{2\pi^2}{3}N(E_F)k_B^2(1+\lambda),\label{eq:5}
\end{equation}
where $N(E_F)$ is the DOS at the Fermi surface and $\lambda$
reflects the coupling strength. The fact that the electron-phonon
coupling strength is weak in present system indicates that the large
value of $\gamma_n$ is not due to the enhanced effective mass but
originates from the high normal state quasiparticle DOS. Comparing
with the $\gamma_n$ obtained for the F-doped LaFeAsO
system\cite{MuG} (about 5-6 mJ/mol-Fe K$^2$), the $N(E_F)$ in
hole-doped FeAs-122 phase may be 5-10 times higher than that in the
electron-doped FeAs-1111 phase. Although the band structure
calculations gave a relatively larger $\gamma_n$ in
FeAs-122\cite{Singh,FangZ,Yildirim,LuZY,Shein} than in FeAS-1111,
the predicted values are however hardly beyond 15 mJ/mol K$^2$ which
is much below the experimental value found here $\gamma_n\approx$
63.3 mJ/mol K$^2$. An appropriate explanation to this large
discrepancy is highly desired from the theoretical side.

\begin{figure}
\includegraphics[width=8.5cm]{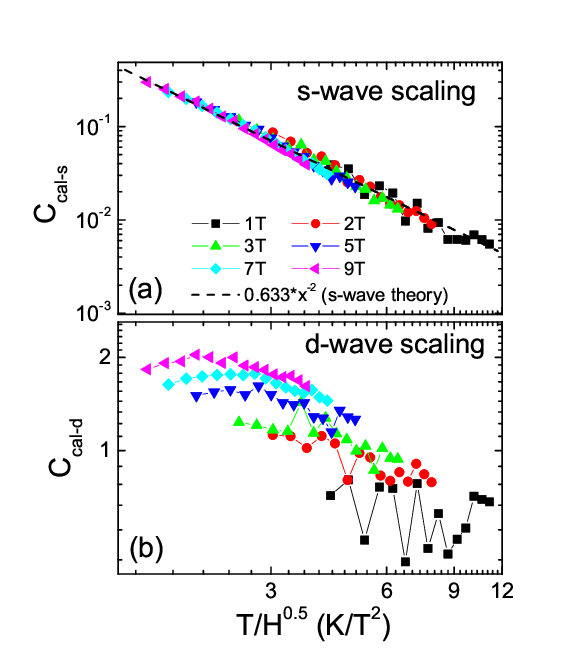}
\caption {(color online) (a) Scaling of the data according to the
s-wave scenario (symbols)
$C_{cal-s}=[(C(H)-C_{Sch}(H))-(C(H=0)-C_{Sch}(H=0))]/T^{3}$ vs.
T$/\sqrt{H}$, the dashed line represents the theoretical expression.
(b) Scaling of the data (symbols) based on d-wave prediction
$C_{cal-d}=[(C(H)-C_{Sch}(H))-(C(H=0)-C_{Sch}(H=0))]/T\sqrt{H}$ vs.
T$/\sqrt{H}$. No good scaling can be found for the d-wave case. }
\label{fig7}
\end{figure}

\begin{figure}
\includegraphics[width=8.5cm]{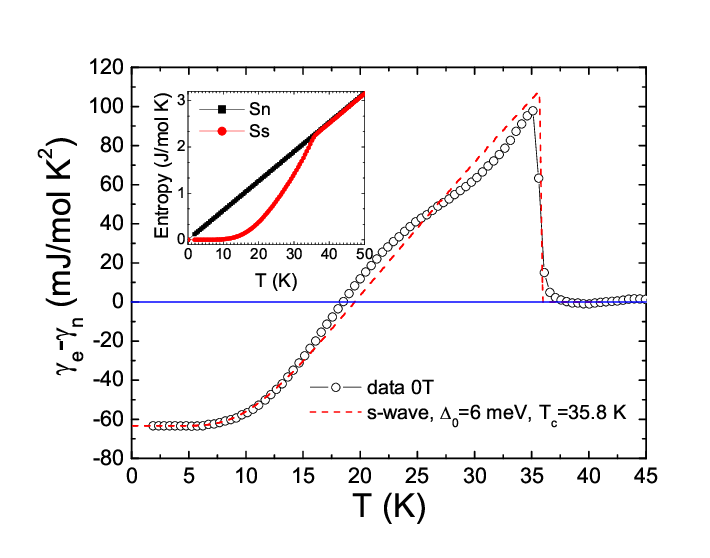}
\caption {(color online) Temperature dependence of the electronic SH
contribution (with the normal state part subtracted) is shown in the
main frame. A sharp SH anomaly can be seen here. A hump is clearly
seen in the middle temperature region. The red dashed line is a
theoretical curve based on the BCS expression with an s-wave gap of
6 meV. The inset shows the entropy of the superconducting state (red
circle symbols) and the normal state (dark square symbols). }
\label{fig8}
\end{figure}

In the FeAs-122 superconductors, it is challenging to measure the
normal state SH below $T_c$ due to the very high $H_{c2}$. In order
to have a comprehensive understanding to the normal state SH, we
have attempted to fit the normal state SH above $T_c$ using a
polynomial function:
\begin{equation}
C_{n}=(\gamma_0 +\gamma_n) T+ \beta_3 T^3+\beta_5 T^5+\beta_7
T^7+\beta_9 T^9+\beta_{11} T^{11},\label{eq:5}
\end{equation}
where we took the values obtained already $\gamma_0 = 7.7$ mJ/mol
K$^2$, $\gamma_n$ = 63.3 mJ/mol K$^2$, and $\beta_3 = 0.473$ mJ/mol
K$^4$.  Other fitting parameters, $\beta_5$, $\beta_7$, $\beta_9$,
and $\beta_{11}$, were left free in the fitting process, yielding
the values of 3.72$\times$ 10$^{-4}$ mJ/mol K$^6$, -5.32$\times$
10$^{-7}$ mJ/mol K$^8$, 2.13$\times$ 10$^{-10}$ mJ/mol K$^{10}$, and
-2.90$\times$ 10$^{-14}$ mJ/mol K$^{12}$, respectively. It's worth
to note that the value of $\beta_5$ is very close to the value of
$\eta$ obtained before. The fitting result of the normal state SH is
displayed by the blue dashed line in the main frame of Fig. 4. The
data after subtracting the normal state SH is presented in the main
frame of Fig. 8. It was found that the entropy-conserving law was
satisfied naturally confirming the validity of our fitting, as shown
in the inset of Fig. 8. A clear flattening of $\gamma_e-\gamma_n$ in
the temperature region below 7 K is observed indicating a fully
gapped superconducting state. Moreover, a hump is clearly seen in
the middle temperature region. We attempted to fit the data using
the BCS formula:
\begin{eqnarray}
\gamma_\mathrm{e}=\frac{4N(E_F)}{k_BT^{3}}\int_{0}^{+\infty}\int_0^{2\pi}\frac{e^{\zeta/k_BT}}{(1+e^{\zeta/k_BT})^{2}}\nonumber\\
\nonumber\\
(\varepsilon^{2}+\Delta^{2}(\theta,T)-\frac{T}{2}\frac{d\Delta^{2}(\theta,T)}{dT})\,d\theta\,d\varepsilon,
\end{eqnarray}
where $\zeta=\sqrt{\varepsilon^2+\Delta^2(T,\theta)}$ and
$\Delta(T,\theta)=\Delta_0(T)$ for the s-wave symmetry. The red
dashed line in the main frame presented the fitting result. One can
see that the fitting curve with a gap value of about 6 meV matched
our data below 13 K perfectly, but failed to describe the hump
feature in the middle temperature region. This hump may be
attributed to the multi-gap effect which seems to appear in the
FeAs-based superconductors, and small components with d-wave gap may
appear in some positions of the Fermi surface. But at this moment we
cannot exclude the possibility that the hump is induced by the
limited uncertainty in getting the normal state phonon contribution.
Nevertheless, the fine fitting in wide temperature region strongly
suggests that the dominant part of the superconducting condensate is
induced by a full gap with the magnitude of about 6 meV. Our results
here seem to be consistent with the ARPES data, both in symmetry and
the small gap\cite{DingH,ZhouXJ,Hasan}. But we have not found a
large gap of 12 meV. This discrepancy may be induced by the
different ways in determining the gap. Future works are certainly
required to reconcile all these distinct results.

\begin{figure}
\includegraphics[width=8cm]{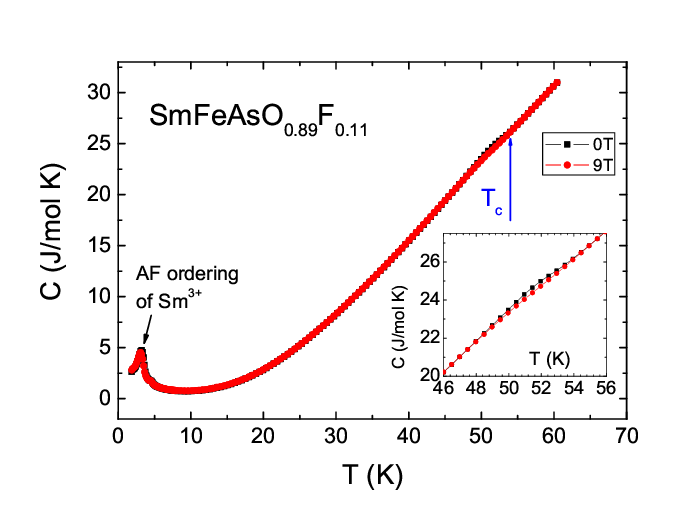}
\caption {(color online) Temperature dependence of SH for the
electron-doped SmFeAsO$_{0.9}$F$_{0.1}$ under 0 T and 9 T is shown
in the main frame. The arrowed blue line denotes the superconducting
transition temperature. A clear peak can be observed at about 3.2 K
under both fields. This peak was observed previously and attributed
to the anti-ferromagnetic ordering of the Sm$^{3+}$ ions. The inset
shows an enlarged view of the main frame near $T_c$. } \label{fig9}
\end{figure}

\begin{figure}
\includegraphics[width=8cm]{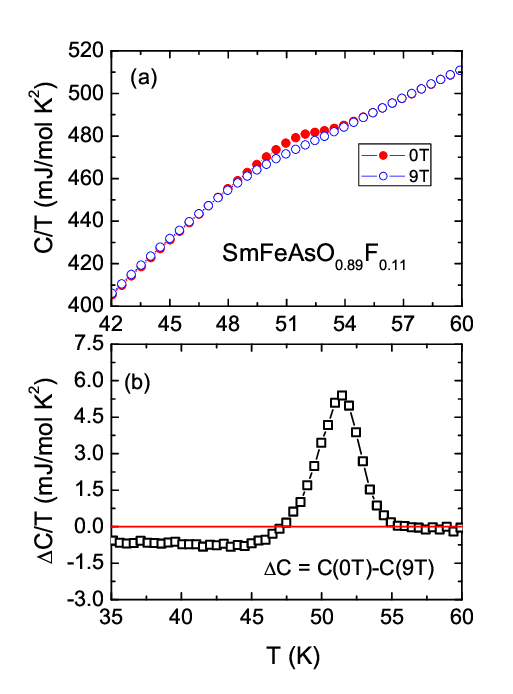}
\caption {(color online) (a) Specific heat data of the sample
SmFeAsO$_{0.9}$F$_{0.1}$ under two different fields 0 T and 9 T,
plotted as C/T vs T close to $T_c$. (b) (C(0 T)-C(9 T))/T vs T for
the same sample. One can see a clear specific heat peak near the
onset temperature of the ac susceptibility curve. } \label{fig10}
\end{figure}

\section{Specific heat of electron-doped $\texttt{SmFeAsO}_{0.9}\texttt{F}_{0.1}$}

We have also investigated the low temperature SH of the
electron-doped SmFeAsO$_{0.9}$F$_{0.1}$ polycrystalline sample with
high quality prepared using a high pressure (HP) synthesis
method\cite{RenZACPL}. In the main frame of Fig. 9, we show the
temperature dependence of the SH in SmFeAsO$_{0.9}$F$_{0.1}$ at 0 T
and 9 T. The clear peak in the low temperature regime (around 3.2 K)
was attributed to the anti-ferromagnetic ordering of Sm$^{3+}$ ions
in this system\cite{SYLi}. The SH anomaly near $T_c$ denoted by the
arrowed blue line is quite unconspicuous, even in the enlarged view
as shown in the inset of Fig. 9, indicating a rather low DOS or
superfluid density in the present system.

The SH data of SmFeAsO$_{0.9}$F$_{0.1}$ near $T_c$ under two
different magnetic fields 0 T and 9 T are plotted as $C/T$ vs $T$ in
Fig. 10(a). The SH anomaly can be observed and it has a width of
about 3 K, which is quite narrow comparing with the $T_c$. This
relatively narrow SH anomaly strongly suggests that the small
magnitude of the SH anomaly is an intrinsic feature, not because the
sample is a polycrystalline one. It is clear that the behavior in
this figure is in sharp contrast with that of the hole-doped
Ba$_{0.6}$K$_{0.4}$Fe$_2$As$_2$ where a sharp and huge peak has been
detected, as shown in the inset of Fig. 4. In Fig. 10(b) we show the
discrepancy of the data in the two curves of Fig. 10(a). A peak due
to superconducting phase transition shows up clearly. The behavior
observed here is quite similar to that reported by other
groups\cite{SYLi} and the noise in our data is much lower. The SH
anomaly $\Delta C/T|_{T_c}$ was estimated to be among 6$\sim$8
mJ/mol K$^2$, being remarkably smaller than that of the FeAs-122
phase. Considering the formula based on the weak-coupling BCS theory
$\Delta C_e/\gamma_nT|_{T_c} $=1.43, we may evaluate the normal
state electron SH coefficient $\gamma_n \approx$ 4$\sim$6 mJ/mol
K$^2$. We have known from Eq. (5) that $\gamma_n$ is proportional to
the normal state quasiparticle DOS $N(E_F)$. So the rather lower
value of $\gamma_n$ in the electron-doped SmFeAsO$_{0.9}$F$_{0.1}$,
compared with that of the hole-doped
Ba$_{0.6}$K$_{0.4}$Fe$_2$As$_2$, clearly illustrates that the normal
state quasiparticle DOS in the electron-doped 1111 phase is
remarkably lower than that in the holed-doped FeAs-122 system.

This large difference may be understood with the complex FS
structures in the two systems. So far the ARPES data are consistent
each other about the shape of the FSs around $\Gamma$
point.\cite{DingH,Zabolotnyy} They also commonly point to a fact of
s-wave gap on the FSs around $\Gamma$, given that the measurements
 about the gap symmetry were not influenced by the surface
reconstruction. However there is a large difference between the
results from different groups about the FSs around the M
point.\cite{DingH,Zabolotnyy,ZhouXJ} It was reported that the FS
near M point has a "propeller" shape with four elliptic blades
surrounding a small circle FS.\cite{Zabolotnyy} How to understand
this complex FS structure is very challenging. Concerning the
pairing symmetry, one possibility is that there are nodes on these
FSs near M point or at the positions of the connecting points of
neighboring FSs. In the Ba$_{0.6}$K$_{0.4}$Fe$_2$As$_2$ samples, the
outer shell of the FSs surrounding $\Gamma$ takes most of the weight
and exhibits an full gap. While in the F-doped LnFeAsO, the FSs near
the M points become dominant and contribute more weight to the
superfluid. This may explain why one can observe the nodal feature
in the electron doped F-doped LnFeAsO samples in specific heat and
point contact tunneling measurements.\cite{MuG,ShanL} Further
momentum resolved experiments on  LnFeAsO single crystals with good
quality are highly desired to resolve this issue.

\section{Concluding remarks}

In summary, the bulk evidence for the dominance of a full-gap
superconducting state with the gap amplitude of about 6 meV was
obtained in the hole-doped Ba$_{0.6}$K$_{0.4}$Fe$_2$As$_2$ through
low temperature SH measurements. An unexpected high SH anomaly
$\Delta C/T|_{T_c}$ and large value of $\gamma_n$ were found,
suggesting a very high normal state quasiparticle DOS. These two
features make the present system very different from the FeAs-1111
phase. We also measured the SH of the high pressure synthesized
electron-doped SmFeAsO$_{0.9}$F$_{0.1}$. It is found that $\Delta
C/T|_{T_c}$ and $\gamma_n$ are only about a tenth of that of
Ba$_{0.6}$K$_{0.4}$Fe$_2$As$_2$. We argue that this difference may
be originated from the complex Fermi surface structures in different
systems.

\begin{acknowledgments}
We acknowledge the fruitful discussions with Tao Xiang, Fuchun
Zhang, Huen-Dung Yang and Dung-Hai Lee. We also thank Prof.
Zhongxian Zhao for providing the high-quality
SmFeAsO$_{0.9}$F$_{0.1}$ samples. This work is supported by the
Natural Science Foundation of China, the Ministry of Science and
Technology of China (973 project No: 2006CB601000, 2006CB921107,
2006CB921802), and Chinese Academy of Sciences (Project ITSNEM).
\end{acknowledgments}

\end{document}